\documentclass[10pt,letterpaper]{article}
\usepackage{opex3}
\usepackage[letterpaper]{geometry}
\usepackage{amsmath,amssymb}
\usepackage{cite}

\usepackage{bm}
\usepackage{graphicx}
\usepackage{color}
\usepackage[active]{srcltx}

\long\def\comment#1{}
\newcommand{\mathcommand}[3][0]{\newcommand{#2}[#1]{\ensuremath{#3}}}
\newcommand{\ts}[2]{{#1}_{\textnormal{#2}}} 

\mathcommand{\radtwo}{\sqrt{2}}
\mathcommand[1]{\oneover}{\frac{1}{#1}}

\newcommand{\be}{\begin{equation}}
\newcommand{\ee}{\end{equation}}

\mathcommand{\eup}{\uparrow}
\mathcommand{\edown}{\downarrow}
\mathcommand{\tup}{\uparrow\downarrow,\Uparrow}
\mathcommand{\tdown}{\uparrow\downarrow,\Downarrow}

\mathcommand[1]{\supop}{\hat{#1}} 
\mathcommand[1]{\hilbert}{\mathfrak{H}_{\text{#1}}} 
\mathcommand[2]{\avehamt}{\ts{\tilde{\mathcal{H}}}{#1}^{(#2)}} 

\mathcommand[1]{\smallket}{|#1\rangle}
\mathcommand[1]{\smallbra}{\langle #1|}
\mathcommand[1]{\bigket}{\bigl|#1\bigr\rangle}
\mathcommand[1]{\bigbra}{\bigl\langle #1\bigr|}
\mathcommand[1]{\biggket}{\biggl|#1\biggr\rangle}
\mathcommand[1]{\biggbra}{\biggl\langle #1\biggr|}
\mathcommand[1]{\ket}{\left| #1 \right\rangle}  
\mathcommand[1]{\lket}{\bigl| #1 \bigr)}  
\mathcommand[1]{\bra}{\left\langle #1 \right|}  
\mathcommand[1]{\lbra}{\bigl( #1 \bigr|}  

\mathcommand[1]{\bbra}{\biggl\langle #1 \biggr|}
\mathcommand[1]{\bket}{\biggl| #1 \biggr\rangle}

\begin{document}

\title{Downconversion quantum interface for a\\
single quantum dot spin and\\
1550-nm single-photon channel}

\author{Jason~S.~Pelc,$^{1,5,*}$ Leo~Yu,$^{1,7}$ Kristiaan~De~Greve,$^{1,6,7}$ Peter~L.~McMahon,$^{1,7}$ Chandra~M.~Natarajan,$^{1,2}$ Vahid Esfandyarpour,$^1$ Sebastian Maier,$^3$ Christian Schneider,$^3$ Martin Kamp,$^3$ Sven H\"ofling,$^{1,3}$ Robert H. Hadfield,$^2$ Alfred~Forchel,$^3$ Yoshihisa~Yamamoto,$^{1,4}$ and M.~M.~Fejer$^1$}

\address{
$^1$ E. L. Ginzton Laboratory, Stanford University, Stanford, CA, USA,\\
$^2$ Scottish Universities Physics Alliance and School of Engineering and Physical Sciences, 
	Heriot-Watt University, Edinburgh EH144AS, UK,\\
$^3$ Technische Physik, Physikalisches Institut,
	Wilhelm Conrad R\"{o}ntgen Research Center for Complex Material Systems,
	Universit\"{a}t W\"{u}rzburg,
	W\"{u}rzburg, Germany,\\
$^4$ National Institute of Informatics,
	Hitotsubashi 2-1-2, Chiyoda-ku,
	Tokyo 101-8403, Japan,\\
$^5$ Now with Hewlett-Packard Laboratories, 1501 Page Mill Rd., Palo Alto, CA, USA.\\
$^6$ Now with Dept. of Physics, Harvard University, Cambridge, MA, USA\\
$^7$ These authors made equal contributions to this work.\\

}
\email{* jason.pelc@hp.com} 


\begin{abstract}
Long-distance quantum communication networks require appropriate interfaces between matter qubit-based nodes and low-loss photonic quantum channels.  We implement a downconversion quantum interface, where the single photons emitted from a semiconductor quantum dot at 910 nm are downconverted to 1560 nm using a fiber-coupled periodically poled lithium niobate waveguide and a 2.2-$\mu$m pulsed pump laser.  The single-photon character of the quantum dot emission is preserved during the downconversion process: we measure a cross-correlation $g^{(2)}(\tau = 0) = 0.17$ using resonant excitation of the quantum dot.  We show that the downconversion interface is fully compatible with coherent optical control of the quantum dot electron spin through the observation of Rabi oscillations in the downconverted photon counts.  These results represent a critical step towards a long-distance hybrid quantum network in which subsystems operating at different wavelengths are connected through quantum frequency conversion devices and 1.5-$\mu$m quantum channels.
\end{abstract}




\section{Introduction}

Long-distance quantum communication networks require appropriate interfaces between matter qubit-based nodes and low-loss photonic quantum channels \cite{kimble_quantum_2008, ritter_elementary_2012, radnaev_quantum_2010}.  Quantum frequency conversion (QFC) \cite{kumar_quantum_1990}, whereby a photonic qubit's carrier frequency is translated while maintaining its quantum state, is well-suited to the task \cite{tanzilli_photonic_2005, ikuta_wide-band_2011}.  Quantum dots have been studied extensively as potential quantum network nodes \cite{press_complete_2008, berezovsky_picosecond_2008}, and upconversion of single photons from quantum dots has been demonstrated \cite{rakher_quantum_2010, rakher_simultaneous_2011}.  However, generation of noise photons, as well as the absence of coherent control in those experiments, have limited the utility of the QFC technique for quantum networks with solid-state matter qubits where resonant excitation is required.  In this paper, we report a low-noise downconversion quantum interface, in which 910-nm single photons from a quantum dot are downconverted to the 1.5-$\mu$m lowest-loss telecom band, showing near-perfect preservation of antibunched photon statistics under coherent quantum dot spin control.  These results represent a critical step towards a long-distance hybrid quantum network \cite{ou_efficient_2008}, in which subsystems operating at different wavelengths are connected through QFC devices and 1.5-$\mu$m quantum channels.

The development of a long-distance hybrid quantum network faces many challenges.  Perhaps foremost among them is an interface between coherently controlled matter qubits (which typically have optical transitions in the visible or near-visible spectral range) and low-loss photonic quantum channels based on transmission through optical fiber.  Quantum frequency conversion enables such an interface.  In this work we demonstrate a QFC scheme whereby single photons emitted from a quantum dot, at a wavelength of 910 nm, are downconverted to the 1.5-$\mu$m lowest-loss telecom band, while the emitted photons are temporally and spectrally shaped by the converter.  QFC has enjoyed renewed interest since the development of upconversion-assisted single-photon detectors at 1.5 $\mu$m: the use of periodically poled lithium niobate (PPLN) waveguides with internal conversion efficiencies exceeding 99.99\% (with extraction efficiencies of $\sim$90\% limited by waveguide propagation losses) for pump powers on the order of 100 mW \cite{langrock_highly_2005, pelc_long-wavelength-pumped_2011}.  There has been recent work on extending QFC for use in downconversion, with experimental efforts focusing on various systems including quantum dots and color centers in diamond \cite{takesue_single-photon_2010, curtz_coherent_2010, pelc_influence_2010, zaske_efficient_2011}, as well as heralded single photons from spontaneous parametric downconversion, in which entanglement properties were shown to be preserved upon downconversion \cite{ikuta_wide-band_2011}.  Our use of a pulsed pump to shape the single-photon wavepacket is a downconversion analog to the technique used in \cite{rakher_simultaneous_2011} to demonstrate fast amplitude modulation of single-photons via upconversion, and in \cite{kuzucu_time-resolved_2008} to achieve high-time-resolution upconversion-based detection.  To the best of our knowledge, there has not yet been a demonstration of a 1.5-$\mu$m downconversion interface for a matter qubit.

\begin{figure}[htb]
\centering
\includegraphics[width=\textwidth]{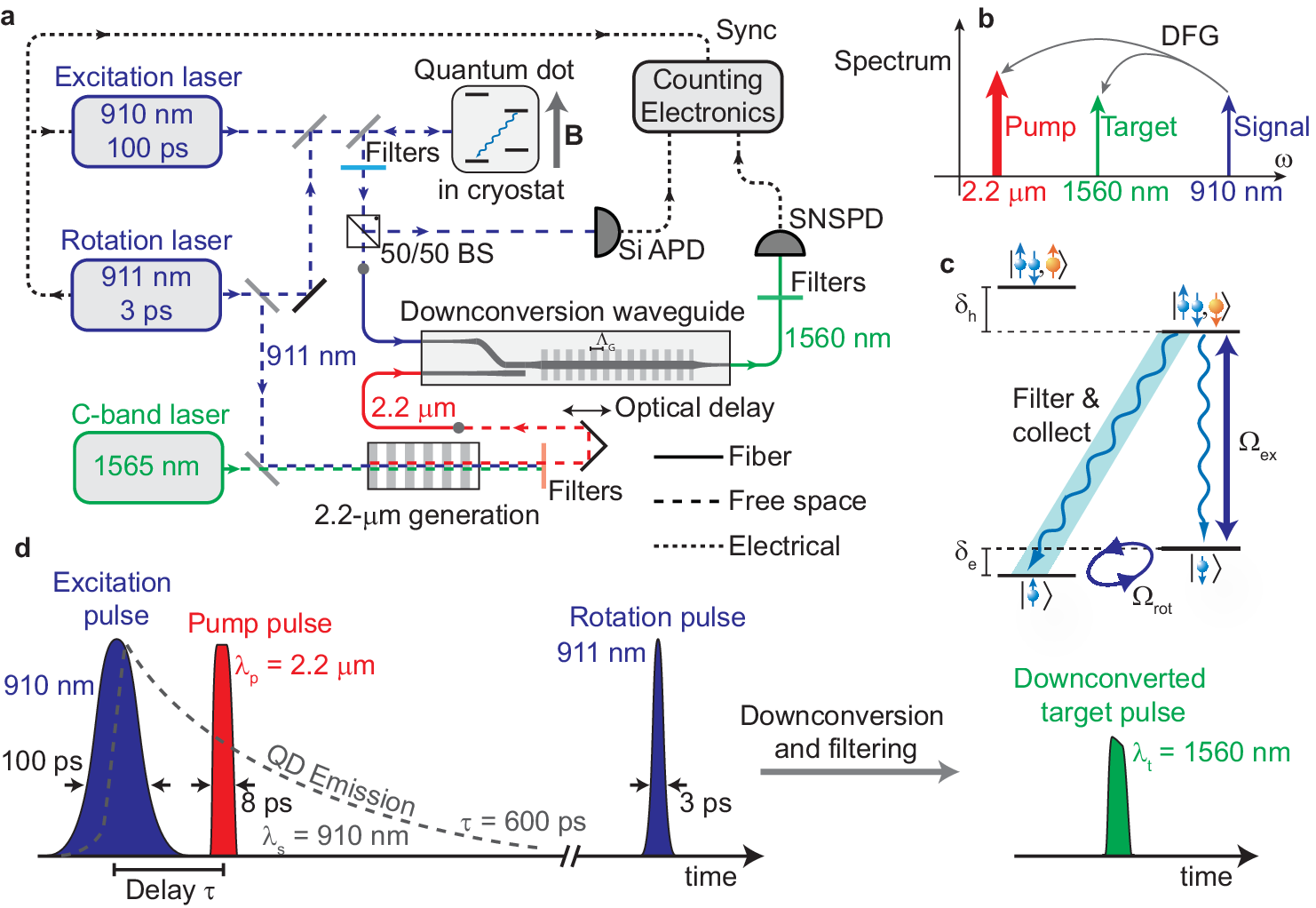}
\caption{(a)-(d), Schematic diagrams of the experiment.  The quantum dot sample was placed in a magnetic field (Voigt geometry) and was resonantly excited on the $\ket{\edown}$--$\ket{\tdown}$ transition (with Rabi frequency $\Omega_\mathrm{ex}$ as in (c) and \cite{press_complete_2008}) with the 910-nm excitation laser. The spin state was controlled with the 911-nm rotation laser (Rabi frequency $\Omega_\mathrm{rot}$), which was synchronized to the excitation laser.  A portion of the rotation laser power was picked off and combined with a strong 1565-nm laser to produce picosecond-pulsed 2.2-$\mu$m radiation used as the pump for the downconversion process.  Single photons emitted via decay from $\ket{\tdown}$ to $\ket{\eup}$ were collected and downconverted in the PPLN waveguide with poling period $\Lambda_G = 21.9$~$\mu$m.  The resulting 1560-nm downconverted photons were detected on a superconducting nanowire single-photon detector (SNSPD).}
\label{fig:1}
\end{figure}

As a matter qubit, we use a single electron in an InAs/GaAs quantum dot (QD) \cite{press_complete_2008, berezovsky_picosecond_2008}.  InAs QDs are among the brightest and fastest known single-photon sources \cite{shields_semiconductor_2007}.  When singly charged, they are also interesting candidates as quantum memories, since ultrafast optical coherent control techniques allow of order $10^5$ single-qubit gate operations before the qubit dephases \cite{press_ultrafast_2010}, and have a level structure shown in Fig.~\ref{fig:1}(c).  Pairs of single photons emitted from the same dot have been shown to be indistinguishable to a high degree \cite{santori_indistinguishable_2002}, and two-photon interference from separate QDs has been demonstrated via precise frequency tuning using either strain \cite{flagg_interference_2010} or electric fields \cite{patel_two-photon_2010}.

There are two telecom bands of interest for long-distance fiber-based quantum networks: the spectral regions near either 1.3 or 1.5 $\mu$m, which have propagation losses of 0.34 or 0.19 dB/km, respectively \cite{takesue_quantum_2007}.  While several QFC experiments have focused on the 1.3-$\mu$m window \cite{rakher_quantum_2010, radnaev_quantum_2010}, an interface to the 1.5-$\mu$m band would enable nearly double the transmission distance of a 1.3-$\mu$m interface.  One difficulty with the construction of a 1.5-$\mu$m interface is that to avoid large noise counts due to spontaneous scattering processes of the strong classical pump, the pump wavelength must be substantially longer than that of the converted signal \cite{pelc_influence_2010, pelc_long-wavelength-pumped_2011}.   

\section{Experiment and results}

\subsection{Frequency conversion setup and characterization}

A system-level schematic of our experimental setup is shown in Fig.~\ref{fig:1}.  Downconversion of the QD single photons at a signal wavelength $\lambda_s  = 910$~nm to a target wavelength $\lambda_t = 1.5$~$\mu$m requires a pump field at $\lambda_p = \left( \lambda_s^{-1} - \lambda_t^{-1} \right)^{-1} =  2.2$~$\mu$m, as shown in Fig.~\ref{fig:1}(b), satisfying the constraint for low noise operation.  To generate the long-wavelength pump, we built a 2.2-$\mu$m picosecond source using a bulk DFG process.  While higher conversion efficiencies could be obtained with a cw pump, a picosecond pulsed pump is used here to enable spectral and temporal shaping of the conversion process, required for future experiments on two-photon interference and entanglement generation \cite{togan_quantum_2010, patel_two-photon_2010}.   We combined the 3-ps 911-nm pulses from a mode-locked Ti:Sapphire laser with pulses from a 2-W telecom-band laser (a tunable diode laser followed by an erbium-doped fiber amplifier) in a bulk MgO:PPLN crystal (length 5 cm, poling period $\Lambda_G = 25.9$~$\mu$m, temperature 177$^\circ$ C).  The crystal length is a factor of 4 longer than the group-velocity walkoff length between the 911-nm pulses and the 1565-nm signal, yielding top-hat-like 2.2-$\mu$m pulses with a full-width at $1/e^2$-intensity maximum of 8.6 ps and a maximum peak power of 40 W.  The MgO:PPLN crystal incorporated multiple QPM grating lengths, allowing us to vary, via choice of grating length, the pulsewidth of the 2.2-$\mu$m pump.

\begin{figure}[t]\begin{center}
\includegraphics[scale=0.9]{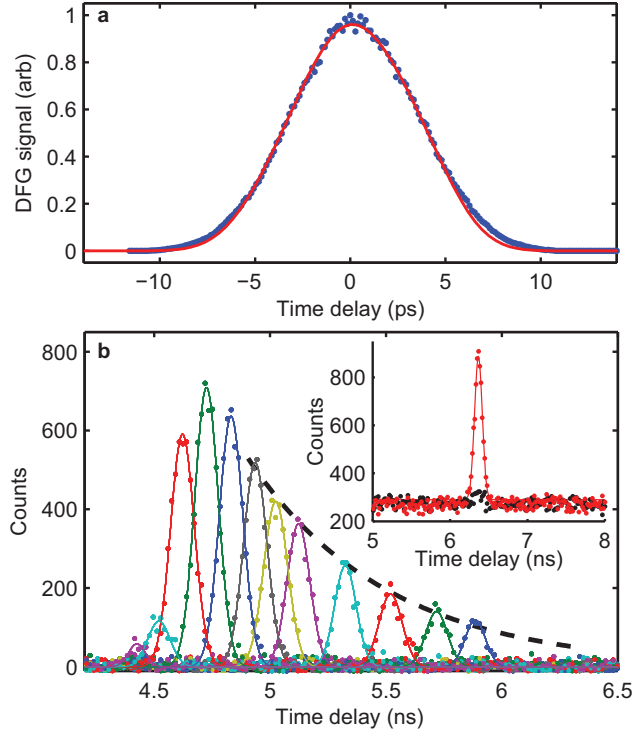}
\end{center}
\caption{Temporal characterization of the downconversion process: (a) DFG cross-correlation for a time delay between 2.2-$\mu$m pump pulses and 3-ps rotation laser pulses in the downconversion waveguide, showing an approximately 10-ps-wide conversion time window, with good agreement to a numerical simulation.
(b), Inset: SNSPD photon-count histogram for an integration time of 300~s showing converted QD single photons (red) and noise when the QD excitation is blocked (black), demonstrating very low leakage.  Main: as the time delay between the excitation and pump pulse is varied, we trace out the spontaneous emission decay curve of the QD; the black dashed curve is an exponential decay with a 600-ps time constant, matching the value observed directly at 910 nm.  For each time delay, we integrated for 300~s.
}
\label{fig:2}
\end{figure}
Single-photon downconversion was done in a periodically poled lithium niobate (PPLN) waveguide fabricated using the reverse-proton-exchange technique \cite{langrock_highly_2005, pelc_long-wavelength-pumped_2011}.  The waveguide chip had a total length of 5.2 cm, had a periodically poled grating of 4-cm length with period $\Lambda_G = 21.9$~$\mu$m.  As shown schematically in Fig.~\ref{fig:1}(a), input coupling was done by pigtailing, using UV-cured optical epoxy, two optical fibers (HI 780 for the signal at $\lambda_s$ and SMF-28 for the pump at $\lambda_p$) held in a v-groove array to two input mode-filter waveguides which were designed to support a single spatial mode at each of these wavelengths.  The pump and signal were combined into the interaction region containing the QPM grating using an integrated WDM coupler \cite{chou} based on a directional coupler in which the pump radiation evanescently couples from one waveguide of the coupler to the other, while the signal (due to its smaller mode size) does not couple.  Following the interaction region, an additional SMF-28 fiber was pigtailed to collect the generated radiation at $\lambda_t$.  We observed optical losses of 1.5 dB (including input-coupling loss and propagation loss) at 910 nm, and the fiber output-coupling loss at 1560~nm was approximately 1 dB.  To measure the temporal resolution of the downconversion system, the fiber-coupled 2.2-$\mu$m radiation was combined in the downconversion waveguide with the 3-ps 911-nm laser pulses, and the DFG radiation at 1.56~$\mu$m was measured as a function of the 2.2-$\mu$m optical delay.  Our results are shown in Fig.~\ref{fig:2}(a) along with a simulation performed using the split-step Fourier method \cite{agrawal_nonlinear_2006}, showing good agreement.  The only free parameters used in the simulation to fit the data in Fig.~\ref{fig:2}(a) were the peak powers of the 2.2-$\mu$m and 911-nm pulses.

The peak internal conversion efficiency of the downconversion waveguide was estimated to be approximately 80\%, limited by the waveguide propagation losses of approximately 0.2 dB/cm, and occurred for peak powers (in the input fiber) of approximately 3 W.  We note that the power for maximum conversion is higher than in previous QFC experiments using similar waveguides due to higher losses observed at $\lambda_p$: we observed a total loss of 7 dB at $\lambda_p$, which includes both input coupling losses and an absorption loss of approximately 1 dB/cm at 2.2-$\mu$m.  The absorption loss results from the tail of an OH$^-$ vibration-libration resonance in LiNbO$_3$, which was enhanced due to the intentional proton doping to form the waveguide \cite{groene_direct_1996}.

\subsection{Quantum dot sample and optical control techniques}

The quantum dot sample used in this work was similar to devices used previously for coherent manipulation and all-optical spin-echo experiments  \cite{press_ultrafast_2010}. It consists of a low density of quantum dots, located at the antinode of an asymmetric, low-$Q$ $(Q \sim 200)$ distributed Bragg reflector (DBR) planar microcavity with center wavelength near 910 nm. The high reflectivity of the bottom-DBR directed the emitted photons towards the collection path, while the relatively low $Q$ allowed for reduction of the optical powers needed for coherent manipulation, and increased control fidelity. A Si $\delta$-doping layer, approximately 10 nm below the quantum dots, charged on average one in three quantum dots; magneto-photoluminescence was used to identify the charged dots. A confocal microscopy setup with high-numerical aperture (NA) collection reduced the effective density of quantum dots within the diffraction-limited laser spot, and spectral inhomogeneity between different dots was used to identify a single one.  The sample was mounted in a superconducting, magnetic cryostat, with base temperature of 1.6 K, and the magnetic field used in the experiments was set to 6~T. An aspheric lens $(\mathrm{NA} = 0.68)$ was mounted inside the cryostat, which improved the photon collection efficiency. A 3-axis slip-stick piezo-actuator setup allowed access to any quantum dot within the sample.  

The QD optical manipulation techniques used in this work were derived from previously reported techniques\cite{press_ultrafast_2010}, with slight modifications. While detuned, 3-ps circularly polarized optical pulses from a modelocked laser (the ``rotation'' laser, detuning $\Delta \approx 1$~nm or 300 GHz) enabled high-fidelity coherent control of the electron spin by means of a Ramsey-interferometer setup, the coherent control of the $\ket{\edown}$--$\ket{\tdown}$ effective two-level system occurs through the use of on-resonant, 100-ps pulses from another modelocked laser, the ``excitation'' laser, which was synchronized to the rotation laser.  The filtering of the spontaneously emitted photon at 910 nm requires a double-grating filtering setup, and accurate cross-polarization before detection on a Si photon counter. 

\subsection{Single-photon downconversion experiments}

For the purpose of downconversion, half of the spontaneously emitted photons are split off and sent into a single-mode fiber, which is connected to the downconversion waveguide.  To filter the downconverted QD photons from residual pump light, parasitic pump second-harmonic radiation, and residual excitation and rotation laser power, we used a fiber-optic circulator and fiber Bragg grating (with 2 nm bandwidth) and a long-pass filter on a Si substrate placed into a fiber U-bench.  The optical filtering setup had approximately 2.4 dB insertion loss.  To detect the downconverted single photons, we used superconducting nanowire single-photon detectors (SNSPDs) which had a 14\% system detection efficiency at 1560 nm with a 40-Hz ungated dark count rate, and were maintained at an operating temperature of 2 K \cite{tanner_enhanced_2010}.  

To coherently generate and convert single photons from the quantum dot, we used the pulse sequence shown in Fig.~\ref{fig:1}(d).  The excitation laser coherently drove the optical transition between states $\ket{\edown}$ and $\ket{\tdown}$, which could then spontaneously decay into either $\ket{\eup}$ or $\ket{\edown}$ \cite{press_complete_2008}.  Photons created upon decay to $\ket{\eup}$ were collected, and after a time delay corresponding to several spontaneous emission lifetimes, a spin-rotation pulse was used to transfer population between $\ket{\eup}$ and $\ket{\edown}$ to avoid trapping the population in $\ket{\eup}$ \cite{press_complete_2008,berezovsky_picosecond_2008}.  The 2.2-$\mu$m pump pulse was used to convert the emitted single photons to 1560 nm, which were then filtered and routed to the SNSPD.  The inset of Fig.~\ref{fig:2}(b) shows a histogram of detected counts on the SNSPD plotted versus the time delay from the rotation laser electrical trigger.  We clearly see a narrow converted pulse (red) when the excitation laser is on, corresponding to downconverted single photons from the QD at $\lambda_t$, and very low noise (black) when the excitation laser is off.  The FWHM temporal width of the conversion feature is 100 ps, due to the timing jitter of the SNSPDs; the SNSPD dark counts give a flat background level.  As the time delay between the excitation and conversion pulse was varied, we traced out the spontaneous emission decay curve of the QD \cite{rakher_simultaneous_2011}.  The dashed black curve is an exponential decay with a 600-ps decay constant, matching closely the value observed directly by time-resolved detection of the emitted photons at $\lambda_s$.  We estimate the rate of incident signal photons at 910 nm in the input fiber of the waveguide to be $6\times10^4$~s$^{-1}$.  Our observed count rates on the SNSPD were approximately 20 signal counts/s, in good agreement with known losses (described above for the waveguide and output filtering), temporal overlap factor of approximately 1.7\% due to the mismatch in durations between the spontaneous emission lifetime of the quantum dot and the 2.2-$\mu$m pump pulsewidth, and the SNSPD detection efficiency of 14\%.

Analyzing the traces in the inset of Fig.~2(b) allows the determination of the approximate noise count rates due to residual leaked pump radiation at 2.2~$\mu$m, control light, and parasitic nonlinear processes in the waveguide.  By fitting the observed noise features to a Gaussian, we find a total noise level (subtracting the dark counts of the SNSPD) of approximately $1.0\pm0.3$ counts/s.  This corresponds to a noise-count probability per 2.2-$\mu$m pump pulse of approximately $(1.3 \pm 0.4)\times 10^{-8}$.  This value could likely be improved with additional filtering, at the expense of system throughput.

\begin{figure}[t]\begin{center}
\includegraphics[scale=0.85]{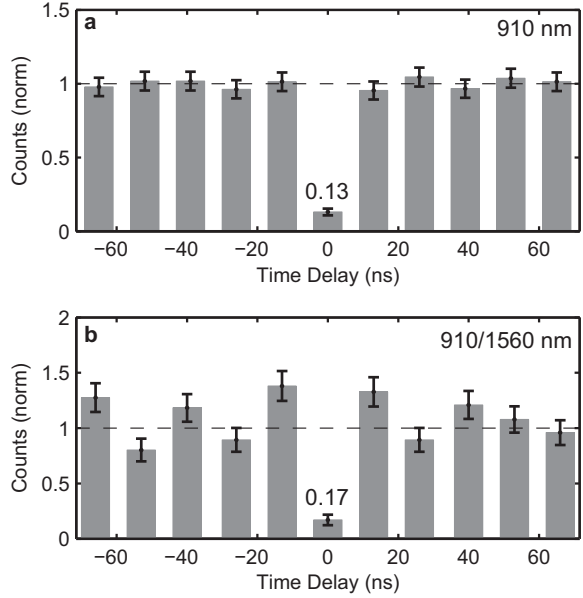}
\end{center}
\caption{Single-photon intensity auto- and cross-correlations, showing strongly antibunched photon statistics.
(a), Measured $g^{(2)}(\tau)$ of the QD emission at 910 nm using two Si APDs and quasi-resonant excitation of the QD, showing a $g^{(2)}(0) = 0.13$.  Each bar represents an integral over a 5-ns time window around each peak.
(b), Normalized cross correlation between nonconverted photons before the waveguide and downconverted photons at 1560 nm, showing $g^{(2)}(0)$ = 0.17.  No background subtraction has been used.  Error bars are one standard deviation assuming Poisson counting statistics.
}
\label{fig:3}
\end{figure}

An important aspect to prove that the quantum frequency downconverter maintains the quantum state of the light emitted from the quantum dot is the investigation of photon statistics.  For this purpose, we began by measuring the second-order autocorrelation function $g^{(2)}(\tau)$ of the quantum dot photons at $\lambda_s$ (with simple modifications to the experimental setup in Fig.~\ref{fig:1}(a)) using a Hanbury Brown and Twiss interferometer and quasi-resonant pulsed excitation \cite{santori_indistinguishable_2002}.  Our results are shown in Fig.~\ref{fig:3}(a), where we see an antibunching dip $g^{(2)}(0) = 0.13$, a signature that we were detecting photons from a single emitter.    Fixing the excitation--conversion delay and pump power, we then measured coincidences between photons reflected off a 50/50 beamsplitter placed before the waveguide, at $\lambda_s$, and the converted photons at $\lambda_t$: our results are shown in Fig.~\ref{fig:3}(b).  We generated counting histograms using a time-correlated single-photon counting system.  The measurement of photon correlations was done using time-tag mode. To suppress the effect of the SNSPD dark counts, the only SNSPD counts that were included in the analysis were those within a 200-ps pulse arrival window.  The fact that the antibunching signature remains (with a zero-delay dip of $g^{(2)}(0) = 0.17$) demonstrates that the downconverter preserves the photon statistics of the input signal upon conversion.  The error bars in Fig.~\ref{fig:3} are calculated from the counting statistics of each peak.  We note that no background subtraction has been applied to the results in Fig.~\ref{fig:3}: in Fig.~\ref{fig:3}(a), the value of $g^{(2)}(0)$ is limited by leakage from the excitation laser, while the value of $g^{(2)}(0)$ in Fig.~\ref{fig:3}(b) has an additional, yet still small, contribution from both the dark counts of the SNSPD and residual leaked light from the 2.2-$\mu$m pump pulse (see Fig.~\ref{fig:2}(b) inset).  To calibrate the reference level $g^{(2)}(\tau) = 1$ in Fig.~\ref{fig:3}(b) 30 accidental correlation peaks ($\tau \neq 0$) were used, giving an expected error in the mean of about 3\%.  We have not measured the  $g^{(2)}(\tau)$ autocorrelation of the generated radiation at the target wavelength because our low count rates would have entailed prohibitively long integration times.  However, due to the very low noise involved in the conversion process and the observations of other authors \cite{ikuta_wide-band_2011, zaske_visible--telecom_2012}, we are confident that the single-photon character of the quantum dot emission is preserved upon downconversion.

\begin{figure}[t]\begin{center}
\includegraphics[scale=0.85]{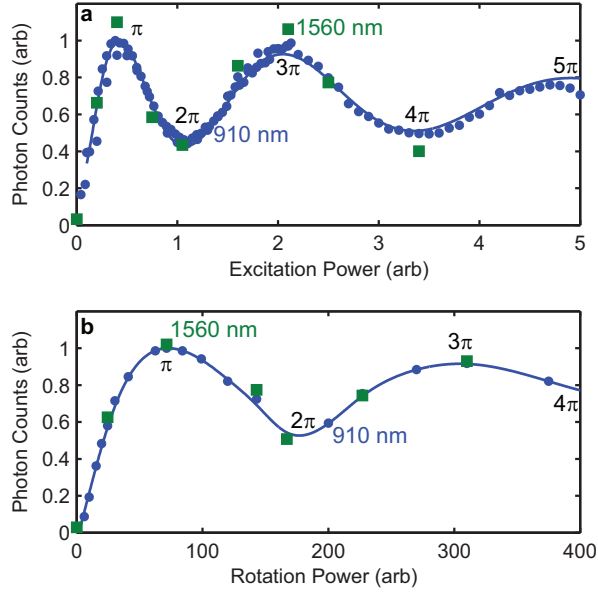}
\end{center}
\caption{
(a), Rabi oscillations, observed as oscillations in the count rate at 910 (blue dots) and 1560 nm (green squares) as the excitation laser power is varied.
(b) shows Rabi oscillations as the QD electron spin state is coherently rotated.  In both plots, the solid blue curves (fit to the 910-nm data) are shown to guide the eye, and the vertical scales of both plots are adjusted to show the agreement between the count rate oscillations at 910 and 1560 nm. The data for the 1560-nm photons were integrated for 300 s per point.
}
\label{fig:4}
\end{figure}

To demonstrate the compatibility of our downconversion interface with coherent control, we performed two Rabi oscillation experiments.  Rabi oscillations indicate that one can perform coherent rotations on a qubit around a single axis.  First, we rotated between $\ket{\edown}$ and $\ket{\tdown}$ using the excitation laser, and second, we rotated between $\ket{\eup}$ and $\ket{\edown}$ using the spin-rotation laser \cite{press_complete_2008}.  The optical-excitation Rabi oscillations, observed directly at 910 nm, are shown as blue dots in Fig.~\ref{fig:4}(a) and the spin-control Rabi oscillations are shown in Fig.~\ref{fig:4}(b), where in both cases the 2.2-$\mu$m pump was off.  Turning on the downconversion pulses at a fixed time delay of 200 ps after the excitation pulse, we observe similar oscillations in the count rates at $\lambda_t = 1.56$~$\mu$m (green squares in Fig.~\ref{fig:4}(a) and (b)), shown on the same vertical scale as the 910-nm data.  As is expected for Rabi oscillations, we observe maxima (minima) in the detected photon counts as the state is rotated through odd (even) integer multiples of $\pi$.  Detailed, quantitative understanding of the control dynamics of the spin system has been reported previously \cite{press_complete_2008, berezovsky_picosecond_2008, press_ultrafast_2010}; we indicate fits to the 910-nm data (blue curves) as a guide to the eye, in good qualitative agreement with the aforementioned models.

\section{Conclusions}

We have demonstrated low-noise downconversion of single photons emitted from a coherently controlled quantum dot to the 1.5-$\mu$m telecommunications band.  The use of picosecond conversion pulses temporally shapes the downconverted emission from the quantum dot.  The resulting time resolution has enabled the generation and verification of telecom-band spin-photon entanglement \cite{togan_quantum_2010, degreve}, which in the QD system requires time resolution well below 50 ps, which would be challenging using presently available single-photon detectors.  

Due to the temporal mismatch between the spontaneous emission lifetime (600 ps) and conversion window (6--10 ps), relatively few of the emitted photons were converted in this experiment.  Similar spectral and temporal shaping, with higher overall efficiency, could be accomplished through the use of a long chirped pump pulse, followed by pulse compression using standard techniques \cite{kielpinski_quantum_2011}.  Specifically, the use of pump pulses with durations matched to the spontaneous emission lifetime would yield an approximately 60-fold improvement in the detected downconverted photon rate, to approximately $1.2\times 10^3$.  Without the losses due to the temporal overlap factor, our external conversion efficiency would be approximately 24\% (including all losses due to the output filters) or 40\% (excluding the filter losses).

In future work, the use of two frequency converters with two pump wavelengths would enable high-fidelity telecom-band interference of photons from quantum dots with different transition frequencies, without resort o temperature or electric-field tuning of the QD transition frequencies \cite{ates_two-photon_2012,takesue_erasing_2008}.  By tuning the pump frequency, a tunable telecom-band single-photon source is created, which has potential uses in wavelength-division multiplexed quantum communications and in quantum metrology.

\section*{Acknowledgements}

This work was supported by the JSPS through its FIRST Program, by NICT, and by the State of Bavaria. JSP acknowledges the support of a Robert N. Noyce Stanford Graduate Fellowship.  JSP and MMF acknowledge additional support from the United States AFOSR.  PLM acknowledges support as a David Cheriton Stanford Graduate Fellow.  KDG acknowledges support as a Herb and Jane Dwight Stanford Graduate Fellow.  CMN acknowledges a SU2P Entrepreneurial Fellowship and RHH acknowledges a Royal Society University Research Fellowship. We gratefully acknowledge Valery Zwiller and Sander Dorenbos at TU Delft, the Netherlands, for providing the superconducting detector samples used.  We thank Chris Phillips and Carsten Langrock for helpful discussions.

\end{document}